\documentclass[twocolumn,authoryear]{autart}    
\usepackage{graphicx}          
\usepackage{amssymb,amsmath}
\usepackage{arydshln}
\usepackage{subfigure}
\usepackage{color}
\usepackage{algorithm}
\usepackage{algpseudocode}
\usepackage{cases}
\usepackage{multirow}
\usepackage[round,sort,authoryear]{natbib}
\usepackage{changes}
\usepackage{cancel}
\usepackage{dsfont}
\usepackage{mathbbol} 
\usepackage{bm}
\usepackage{stfloats}
\usepackage{setspace}
\usepackage{enumerate}
\usepackage{ulem}
 \usepackage{booktabs}

\newtheorem{lemma}{\bf Lemma}
\newtheorem{remark}{\bf Remark}
\newtheorem{corollary}{\bf Corollary}
\newtheorem{proposition}{\bf Proposition}
\newtheorem{definition}{\bf Definition}
\newtheorem{ft}{\bf Fact}
\newtheorem{problem}{\bf Problem}

\newcommand{\hd}[1]{\textcolor{black}{#1}}
\newcommand{\hdb}[1]{\textcolor{black}{#1}}

\newcommand{\hwz}[1]{\textcolor{black}{#1}} 
\newcommand{\pl}[1]{\textcolor{black}{#1}}
\newcommand{\hwzh}[1]{\textcolor{black}{#1}} 
\newcommand{\hdba}[1]{\textcolor{black}{#1}}
\newcommand{\hdbai}[1]{\textcolor{black}{#1}}

\begin{document}

\begin{frontmatter}

\title{Restricting Voltage Deviation of DC Microgrids with Critical and Ordinary Nodes: \hwz{A Generalized Consensus Approach} \thanksref{footnoteinfo}} 

\thanks[footnoteinfo]{This work was supported by Shenzhen Science and Technology Program under grants GXWD20231129102406001 and JCYJ20220818102416036,  the Key Scientific Research Projects of Higher Education Institutions in Henan Province under grant 24A120011, and the National Natural Science Foundation of China under grant 62303133. 
\newline
* Corresponding author.}

\author[YRCTI]{Handong~Bai}\ead{baihandong@yrcti.edu.cn},    
\author[HIT]{Peng~Li}\ead{lipeng2020@hit.edu.cn},
\author[HIT]{Hongwei~Zhang}*\ead{hwzhang@hit.edu.cn}  

\address[YRCTI]{School of Electrical Engineering, Yellow River Conservancy Technical Institute, Kaifeng, Henan 475004, P.R. China}  
\address[HIT] {School of Intelligence Science and Engineering, Harbin Institute of Technology, Shenzhen, Guangdong 518055, P.R. China}

\begin{keyword}                           
compromised control; current sharing; DC microgrid;  Kron reduction; voltage deviation.
\end{keyword}                             

\begin{abstract}                          

Restricting bus voltage deviation is crucial for normal operation of multi-bus DC microgrids, yet it has received insufficient attention due to the conflict between two main control objectives in DC microgrids, i.e., voltage regulation and current sharing.
By revealing a necessary and sufficient condition for achieving these two objectives,
this paper proposes a novel consensus-based current sharing control law that can achieve the compromised control objective, balancing both current sharing and voltage deviation restriction. Additionally, we examine the effectiveness of the proposed control scheme for DC Microgrids that include both critical nodes and ordinary nodes, where there is a simultaneous requirement for voltage deviation limits on critical nodes and accurate current sharing among ordinary nodes. 
Theoretical results are verified by simulations, and the effectiveness in handling plug-and-play operations of distributed generators is also illustrated.
\end{abstract}

\end{frontmatter}

\section{Introduction}

Distributed control of DC microgrids has received extensive attention in both control and power communities 
\citep{Persis2018power,Che2014DCeconomic,dragivcevic2015dc,Liuxk2023auto_stabilityDCMG}.
Two common  objectives for DC microgrids are current sharing and voltage regulation \citep{dragivcevic2015dc}.
The former addresses the problem of sharing the total load current among distributed generators (DGs) in proportion to their current ratings, and
the latter aims to regulate each bus voltage to its rated value.
However, there exists a conflict between current sharing and voltage regulation \citep{Handong2022auto,HanR2019}.
Most existing control methods prioritize current sharing over voltage regulation, and can only achieve voltage balancing  that  regulates the steady-state average or weighted-average voltage over all buses at a rated value \citep{nahata2020consensus,Trip2019average,cucuzzella2018robust,shafiee2014DCmicrogrid,Tucci2018,Liuzj2024TPS}.

For normal operation of loads, voltage deviations of buses from the rated value should be kept within an admissible range, \hdbai{e.g.,} $5\%$ of the rated voltage \citep{Prabhakaran2018TPE_nonlineardroop,NasirianFL2015-TPE}.
Even though accurate current sharing and voltage balancing are achieved,
voltage deviations exceeding admissible   ranges may cause abnormal operation or damage to certain loads.
Thus, bus voltage deviations should be strictly restricted,  
especially for voltage-sensitive loads.

The issue of voltage deviation restriction was concerned in \cite{HanR2019}, and a compromised control method was proposed to achieve a trade-off between current sharing and voltage regulation.
\cite{LDing2018} proposed a distributed optimal control strategy to achieve a compromised objective between accurate current sharing and voltage regulation.
In \cite{cucuzzella2018robust}, a manifold was designed to take both current sharing and voltage regulation into consideration, in which voltage deviation can be regulated. Yet, in these studies, the mechanism of interaction between current sharing and voltage regulation is not disclosed, and thus it is hard to quantitatively adjust the accuracy of current sharing and voltage deviation regulation.
Voltage deviation regulation was further investigated in \citep{Handong2022auto}, where a distributed compromised controller  was proposed to achieve a trade-off between accurate current sharing and reference voltage \hdbai{consensus}. 
However, it cannot regulate each bus voltage to its rated value simultaneously, and thus fails  when strict voltage deviation is required.

Furthermore, in the existing literature \hdbai{on} distributed control of DC microgrids, most  works treat all nodes equally.
However,
different types of loads have various tolerance to voltage deviations.
For example, data centers are voltage-sensitive loads that often require a very strict voltage deviation of less than $2\%$ of the rated voltage \citep{Pratt2007Vdeviation}, while voltage-insensitive loads like residential facilities can even tolerate a 
voltage deviation of $8\%$ of the rated voltage.
Therefore, \pl{
in DC microgrids, nodes should be classified regarding voltage deviation and current sharing.}
To the best of our knowledge, such control issues have not been investigated so far.
Therefore, this paper aims to investigate DC microgrids consisting of critical nodes (i.e., voltage-sensitive nodes) and ordinary nodes (i.e., voltage-insensitive ones), and design distributed control laws that can simultaneously strictly restrict voltage deviations for critical nodes and achieve accurate current sharing for ordinary nodes.
The main contributions of this paper are as follows: 
\begin{enumerate}[(a)]
  \item Control objectives of accurate current sharing, accurate voltage consensus, and voltage-current compromise are uniformly characterized by voltage and current deviation ratios, facilitating the controller design. In addition, necessary and sufficient condition for the existence of the conflict between accurate current sharing and voltage consensus is derived.
  \item Kron reduction 
  is  applied to reduce the topology of the communication network for the first time. This allows us to control DC microgrids with two classes of nodes, achieving  distinct control objectives.
  \item A generalized consensus control framework is proposed, and voltage deviation restriction of critical nodes and accurate current sharing of ordinary nodes are achieved simultaneously.
  The monotonicity of voltage and current deviation ratios is guaranteed, which is of practical significance for designing the trade-off factor.
\end{enumerate}

\section{Preliminaries and problem formulation}
\subsection{Notations}
The identity matrix is denoted by $ \bm{E}$.
A zero vector or matrix with appropriate dimensions is denoted by $\mathbb 0$ .
Let $\mathbb 1_n=[1,\cdots,1]^T\in\mathbb R^n$.
A diagonal matrix  with $g_i$ being the $i$-th  \hdbai{diagonal} entry is  $diag(g_1, g_2, \cdots, g_n)$.
A matrix $\bm A=\left[a_{ij}\right]\in \mathbb{R}^{n\times n}$ is nonnegative (positive), denoted by $\bm A \geq \mathbb 0$ ($\bm A > \mathbb 0$), if  $a_{ij}\geq 0$ ($a_{ij}> 0$), $\forall i, j$. \hwz{Matrix $|\bm A|$ is the element-wise absolute value of $\bm A$. }

\subsection{Graph theory}
The communication network of a DC microgrid is modeled as a graph $\mathcal{G} =(\mathcal{V},\mathcal{E})$, where $\mathcal{V}=\left\{ v_1,\cdots, v_N \right\}$ denotes the set of nodes, i.e., DGs or buses, and $\mathcal{E} \subseteq \mathcal{V} \times \mathcal{V} $ denotes the set of edges, i.e., communication links.
The adjacency matrix of a graph is $\mathcal{A}=[a_{ij}]\in {\mathbb{R}}^{N\times N}$, where $a_{ij}$ is the weight of edge $(v_j, v_i)$, and $a_{ij}>0$ if $(v_j, v_i) \in \mathcal{E}$; otherwise $a_{ij}=0$. The neighbor set of node $i$ is $\mathcal{N}_i=\{j |  (v_j, v_i) \in \mathcal{E}\}$.  
The Laplacian matrix $\mathcal{L}=[l_{ij}]\in\mathbb{R}^{N\times N}$ of $\mathcal{G}$ is defined as $l_{ii}=\sum\nolimits_{j=1}^{N}{a_{ij}}$ and $l_{ij}=-a_{ij}$ if $i\ne j$.
In this paper,  $\mathcal{G}$ is undirected and connected \citep{FLewis2014Book-distributed}.

\subsection{DC microgrid modeling}
Consider a DC microgrid consisting of $N$ ($N>1$) nodes \hwzh{ and $M$ power lines}, whose electrical network has been reduced by Kron reduction \citep{dorfler2013TCSI}.
\begin{figure}[htbp]
  \centering
  \includegraphics[width=7.5cm]{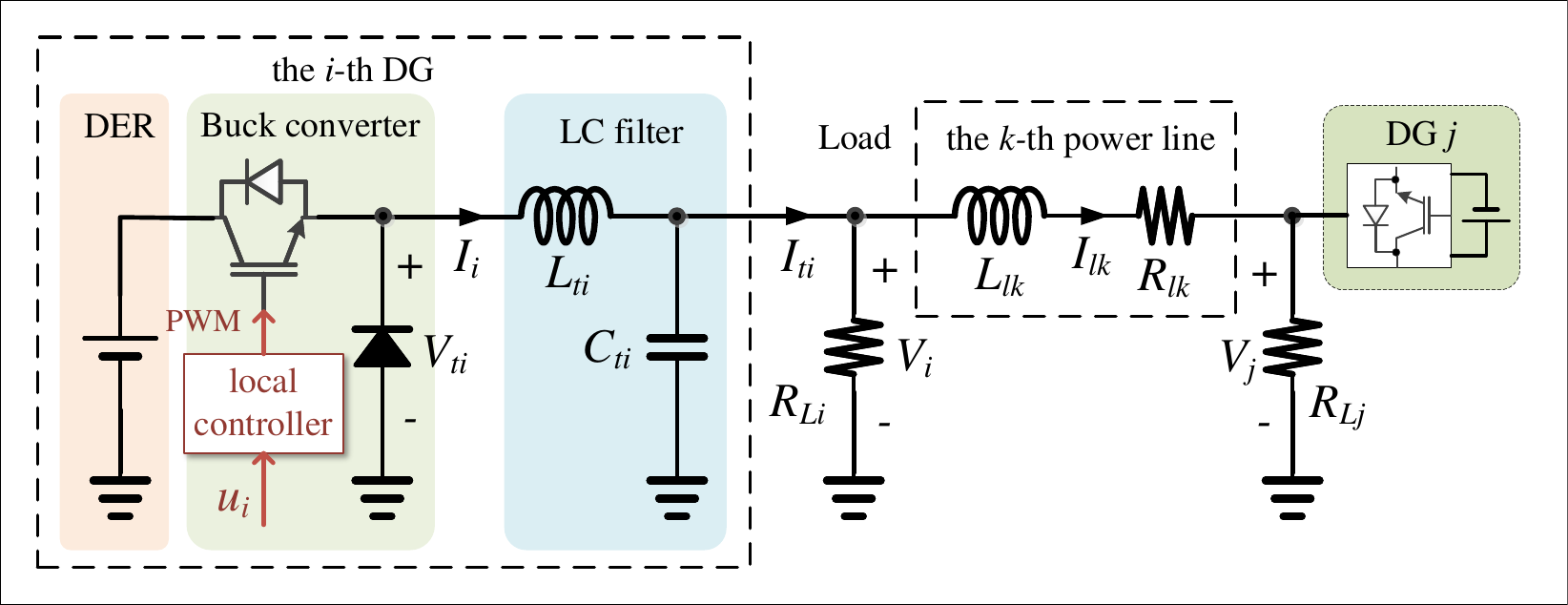}\\
  \caption{Electrical scheme of the $i$-th DG \citep{Trip2019average}} \label{fig:scheme_DG}
\end{figure}
A general electrical scheme of the $i$-th DG is shown in Fig. \ref{fig:scheme_DG}, where $V_{t_i}$ denotes the output voltage of the Buck converter of the $i$-th DG, $L_{t_i}$ and $C_{t_i}$ denote the inductance and capacitance of the LC filter respectively.
Following standard practices \citep{SchifferJ2016,nahata2020}, we have $V_{t_i}=u_i$,
where $V_{t_i}$ is the output voltage of the $i$-th DG converter, and $u_i$ is the associated control signal for this converter.
Let $\bm V_t=[V_{t_1},\cdots,V_{t_N}]^T$ and $\bm u=[u_1,\cdots,u_N]^T$. Then
\begin{equation}\label{e_source}
  \bm V_t=\bm u.
\end{equation}
The dynamics of power lines can be modeled as 
\begin{equation}\label{e_grid}
  \bm L_l\dot{\bm I}_l = \bm B^T\bm V-\bm R_l\bm I_l,
\end{equation}
where $\bm V=[V_1,\cdots,V_N]^T$ and $\bm I_l=[I_{l_1},\cdots,I_{l_M}]^T$ denote bus voltages and the currents of power lines, respectively;
$\bm L_l=diag(L_{l_1},\cdots,L_{l_M})$ and $\bm R_l=diag(R_{l_1},\cdots,R_{l_M})$ denote inductances and resistances of power lines, respectively;
$\bm B=[B_{ik}]\in\mathbb R^{N\times M}$ denotes the incidence matrix of electrical topology of the DC microgrid, in which $B_{ik}=+1$ (\hwzh{$-1$}) if the $i$-th bus is the source (\hwzh{sink}) of the $k$-th power line; otherwise, $B_{ik}=0$.

This paper considers loads of constant impedances with admittances
$\bm Y_L=diag(Y_{L_1},\cdots,Y_{L_N})$,
where $Y_{L_i}=\frac{1}{R_{L_i}}$, and $Y_{L_i}, R_{L_i}$ are conductance and resistance of the $i$-th load, respectively.
According to Kirchhoff's voltage and current laws, the dynamics of the output LC filters of the converters can be modeled as
\begin{subequations}\label{e_LC}
\begin{align}
  \bm L_t\dot{\bm I} &= \bm V_t-\bm V \label{e_LC-a}\\
  \bm C_t\dot{\bm V} &= \bm I-\bm Y_L\bm V-\bm B\bm I_l \label{e_LC-b},
\end{align}
\end{subequations}
where $\bm I=[I_1,\cdots,I_N]^T$ denotes the output currents of the DG converters; $\bm L_t=diag(L_{t_1},\cdots,L_{t_N})$ and $\bm C_t=diag(C_{t_1},\cdots,C_{t_N})$ denote inductances and capacitances of the LC filters, respectively.

\subsection{Conflict of voltage consensus and current sharing}
\hwzh{The objectives of } proportional current sharing among DGs and maintaining all bus voltages at the rated value cannot be achieved simultaneously \citep{Liuxk2023auto_stabilityDCMG}.
This conflict will be rigorously analyzed in the sequel.
The following definitions and fact are instrumental.
\begin{definition}\citep{Handong2022auto}\label{def_devi_V}
In a DC microgrid, the voltage deviation ratio and the current deviation ratio for the $i$-th DG are defined respectively as
\begin{equation*}
  \Delta^V_i=\frac{V_i-V_{rat}}{V_{rat}}\quad and\quad \Delta^I_i=\frac{I^{pu}_i-I^{pu}_{avg}}{I^{pu}_{avg}},
\end{equation*}
where $V_{rat}$ is the rated bus voltage of the DC microgrid, $I^{pu}_i={I_i}/{I^*_i}$ is the output per-unit current of the DG at node $i$ with $I^*_i>0$ being the current capacity of the $i$-th DG, and $I^{pu}_{avg}=\sum_{i=1}^NI^{pu}_i/N$.
Furthermore, define the maximum voltage deviation ratio and the maximum current deviation ratio of the DC microgrid as $\Delta^V_{max} = \left\|\bm\Delta^V\right\|_\infty$ and $\Delta^I_{max} = \left\|\bm\Delta^I\right\|_\infty$, respectively,
where $\bm \Delta^V=[\Delta^V_1,\cdots,\Delta^V_N]^T, \bm \Delta^I=[\Delta^I_1,\cdots,\Delta^I_N]^T$.
\end{definition}
Based on the above notions, accurate current sharing and voltage consensus are further defined.
\begin{definition}
[Accurate current sharing]\label{def_aI}
In a DC microgrid, accurate current sharing among DGs is achieved if $\Delta^I_{max}=0$, i.e.,
$\frac{I_i}{I_i^*}=\cdots
=\frac{I_N}{I_N^*}.$
\end{definition}
\begin{definition}
[Accurate voltage consensus]\label{def_aV}
In a DC microgrid, accurate voltage consensus among buses is achieved if $\Delta^V_{max}=0$, i.e.,
$V_i=\cdots=V_N=V_{rat}.$
\end{definition}
The following Fact \hwzh{shows when these two objectives conflict with each other}.
\begin{ft}\label{ft_conflict}
Consider a DC microgrid described by \eqref{e_source}, \eqref{e_grid} and \eqref{e_LC}. In steady state, 
accurate current sharing and \hwzh{accurate} voltage consensus can be achieved simultaneously if and only if  $Y_{L_i}>0$ for all $i\in \{1,2,\dots,N\}$ and
      \begin{equation}\label{e_noconflict}
        \frac{I^*_1}{Y_{L_1}}=\frac{I^*_2}{Y_{L_2}}=\cdots=\frac{I^*_N}{Y_{L_N}}.
      \end{equation}
\end{ft}
\begin{remark}
Equation \eqref{e_noconflict} implies \hwzh{that} the rated output current of each node matches its load. For this case, when the DG of each node injects its rated current, i.e.,$I_i=I_i^*$, the steady state bus voltages $V_i=I_i/Y_{L_i}=I_iR_{L_i}$ of all nodes are equal. Then there is no voltage difference between buses and no current flows between them.
\end{remark}
When  \eqref{e_noconflict}  does not hold, \hwzh{a conflict emerges and} a concept of voltage-current compromise is introduced.

\begin{definition}[Voltage-current compromise]
Let $\bar\Delta^V$ be the maximum voltage deviation ratio when accurate current sharing is achieved, and  $\bar\Delta^I$ be the maximum current deviation ratio when accurate voltage consensus is achieved.
Then, voltage-current compromise is established if $\Delta^V_{max}<\bar\Delta^V$ and $\Delta^I_{max}<\bar\Delta^I$.
\end{definition}

\subsection{Problem formulation}
A practical DC microgrid usually consists of two classes of loads, i.e., voltage-sensitive and voltage-insensitive loads. While it is necessary to restrict bus voltage deviations for voltage-sensitive loads within their normal operational ranges, voltage regulation of voltage-insensitive loads is not a major concern. Thus, we shall call  voltage-sensitive nodes as \textit{critical nodes} and  voltage-insensitive nodes as \textit{ordinary nodes}.

Consider a DC microgrid with $N$ nodes and let $\mathcal{M}=\{1,\cdots,N\}$. Without loss of generality, let  $\hd{\mathcal{M}_c}=\{1,\cdots,m\}$ be the set of critical nodes, and $\mathcal{M}_o=\{m+1,\cdots,N\}$ be the set of ordinary nodes.
The main problem can then be formulated as follows.
\begin{problem}\label{plm_1}
Consider a DC microgrid described by \eqref{e_source}, \eqref{e_grid} and \eqref{e_LC}, which consists of $m$ critical nodes and $N-m$ ordinary nodes.
  Let   $\Gamma_V\in\left[0,\bar{\Delta}^V_c\right]$ be a user defined voltage deviation index, where $\bar{\Delta}^V_c$ is the maximum voltage deviation ratio of critical nodes when their accurate current sharing is achieved. Let $V_{rat}$ be the desired rated value for critical nodes. Design a distributed controller in the form $ u_i(\theta, \omega, I_i, V_i, \phi_j|j\in \mathcal{N}_i)$, where $\theta$ and $\omega$ are design parameters and $\phi_j$ stands for some neighborhood variables, such that the following objectives can be obtained.
 \begin{enumerate}[(a)]
      \item \label{obj-a}
    For critical nodes, their maximum voltage deviation ratio is restricted by $\Gamma_V$ and their average voltage is regulated to the rated value, i.e., 
    \vspace{-10pt}
    \begin{equation} \label{problem-a}
    \mathop{max}\limits_{i\in \mathcal{M}_c}{\Delta_i^V}\le\Gamma_V, ~~ \frac{1}{m}\sum_i^m V_i =V_{rat}, ~i\in \mathcal{M}_c
    \end{equation}
     \vspace{-10pt}
     \item \label{obj-b}
     For critical nodes, their voltage deviation ratios $\Delta^V_i$ can be monotonically adjusted by $\theta$.
     \item \label{obj-c}
     For ordinary nodes, accurate current sharing is achieved and the per-unit current $I_i^{pu}={I_i}/{I_i^*}$ can be monotonically adjusted by the design parameter $\omega$ , i.e.,
    $
          {I_i}/{I_i^*}={I_j}/{I_j^*}=\alpha(\omega), ~ \forall i, j\in \mathcal{M}_o,
     $
     where the function $0\leq \alpha(\omega)\leq 1$ is a monotonic function of $\omega$.
 \end{enumerate}
\end{problem}
\begin{remark}
    Problem \ref{plm_1} focuses on the case where $\Gamma_V\in\left[0,\bar{\Delta}^V_c\right]$. Smaller $\Gamma_V$ implies more strict voltage deviation at critical nodes. For the extreme case of $\Gamma_V=0$, voltages of all critical nodes should \hwzh{achieve} consensus. 
    The ability to monotonically adjust $\Delta^V_i, i\in\mathcal{M}_c$ along a single parameter $\theta$ is practically important, as it provides practitioners a convenient way to adjust the voltage deviation of critical nodes. For a very conservative value $\Gamma_V>\bar{\Delta}^V_c$, one only needs to control all critical nodes to achieve accurate current sharing, and then the Objective (\ref{obj-a}) will be automatically achieved. In this case, no voltage-current compromise is necessary and many \hdbai{existing} works can deal with it. The Objective (\ref{obj-c}) implies that the \hwzh{portion of } contribution of ordinary nodes to the microgrid can be monotonically tuned by the parameter $\omega$. 
\end{remark}
\section{Compromised controller design}\label{sec_compromisedctrl}
\subsection{Average voltage estimation for critical nodes}
To address Problem \ref{plm_1}, we first design \hwzh{ an } average voltage estimator of critical nodes.
Inspired by the average consensus algorithm proposed in \cite{Freeman2019MCS}, an average voltage observer for microgrid can be put as  
\begin{equation}\label{e_par_avg}
  \left[
     \begin{array}{c}
       \dot{\bar{\bm V}}_c \\
       \dot{\bar{\bm V}}_o \\
     \end{array}
   \right]=\left[\begin{array}{c}
                     \dot{\bm V}_c \\
                     \dot{\bm V}_o \\
                 \end{array}
           \right]-\left[
              \begin{array}{cc}
                \mathcal L_{11} & \mathcal L_{12} \\
                \mathcal L_{21} & \mathcal L_{22} \\
              \end{array}
            \right]\left[
                     \begin{array}{c}
                       \bar{\bm V}_c \\
                       \bar{\bm V}_o \\
                     \end{array}
                   \right],
\end{equation}
where $\bar{\bm V}_c$ and ${\bm V}_c$ are the average voltage estimates and measured voltages of all critical nodes, and $\bar{\bm V}_o$ and ${\bm V}_o$ are the average voltage estimates and measured voltages of all ordinary nodes.
Manipulating \eqref{e_par_avg} yields that
    $\dot{\bar{\bm V}}_c=\dot{\bm V}_c-\mathcal{L}_{22}|\mathcal{L}\bar{\bm V}_c+\mathcal{L}_{12}\mathcal{L}_{22}^{-1}\left(\dot{\bar{\bm V}}_o-\dot{\bm V}_o\right),$
with $\mathcal L_{22}|\mathcal L=\mathcal L_{11}-\mathcal L_{12}\mathcal L_{22}^{-1}\mathcal L_{21}$. Obviously, if $\dot{\bar{\bm V}}_o -\dot{\bm V}_o=\bm 0$, we have 
   $ \dot{\bar{\bm V}}_c=\dot{\bm V}_c-\mathcal{L}_{22}|\mathcal{L}\bar{\bm V}_c.$
Accordingly, \hdba{we design an estimator $\tilde{V}_i$ of average voltage among critical  nodes as} 
\begin{align}
    \dot{\tilde V}_i &=\dot{V}_i+\sum^N_{j=1}a_{ij}\left(\tilde V_j-\tilde V_i\right),   \quad \forall i\in \hd{\mathcal{M}_c} \label{e_cnode_obsv}\\
    \tilde V_k &=\frac{\sum^N_{j=1}a_{kj}\tilde V_j}{\sum^N_{j=1}a_{kj}},  \quad \forall k\in \hd{\mathcal{M}_o} \label{e_cnode_obsv-b}
\end{align}
where 
$a_{ij}, a_{kj}$ are the elements of the adjacency matrix $\mathcal A$ of the communication graph.

The above procedure shows how to  eliminate ordinary nodes from the communication graph using  Kron reduction technique \citep{CALISKAN20142586,dorfler2018ProcIEEE-electr}. This is crucial because it decouples two classes of nodes in a microgrid, thereby simplifying the controller design to meet various objectives.

\begin{proposition}\label{prop_Vcri_obser}
Consider the DC microgrid \eqref{e_source}, \eqref{e_grid} and \eqref{e_LC} consisting of  critical nodes $i\in\mathcal{M}_c$ and  ordinary nodes $k\in\mathcal{M}_o$. In steady state, the estimation governed by \eqref{e_cnode_obsv} and \eqref{e_cnode_obsv-b} converges to the average voltage among the critical nodes, i.e.,
   $\lim_{t\to\infty}\tilde{V}_i=\frac{1}{m}\sum^m_{j=1}V_j, \quad \forall i\in\mathcal{M}.$
\end{proposition}

\subsection{Distributed control law design}
A  distributed secondary control law has been broadly used for accurate current sharing 
with a generic form:
\begin{equation}\label{e_accurateI_consensus}   \dot{x}_i=\sum^N_{j=1}a_{ij}\left(\frac{I_j}{I_j^*}-\frac{I_i}{I_i^*}\right), \forall i\in\mathcal{M},
\end{equation}
where \hwzh{$x_i$} denotes different notions in different works, such as voltage correction term in \citet{NasirianFL2015-TPE}, additional state variable in \citet{nahata2020consensus} and voltage deviation between voltage reference and node voltage in \citet{Tucci2018}.
In steady state, accurate current sharing is achieved.
However, the distinct objectives for critical and ordinary nodes, as addressed in Problem \ref{plm_1}, \hdbai{have} not been simultaneously considered in all aforementioned works. Inspired by \eqref{e_accurateI_consensus}, we shall propose a general consensus-based control framework that can simultaneously address \hdbai{separate} objectives of Problem \ref{plm_1}.

Before designing the compromised control laws for critical nodes, we need the following preliminary result.
Let $\bm Y=\bm B\bm R_l^{-1}\bm B^T$ and
  $\bar{\bm Y}=\bm Y+\bm Y_L=\left[
              \begin{array}{cc}
                \bar{\bm Y}_{11} &\bm Y_{12} \\
               \bm Y_{21} & \bar{\bm Y}_{22} \\
              \end{array}
            \right]$,
where $\bar{\bm Y}_{11}\in\mathbb R^{m\times m}, \bm Y_{12}\in\mathbb R^{m\times (N-m)}, \bm Y_{21}\in\mathbb R^{(N-m)\times m}, \bar{\bm Y}_{22}\in\mathbb R^{(N-m)\times (N-m)}$.
\begin{lemma}\label{ft_conflict_2}
Consider the DC microgrid \eqref{e_source}, \eqref{e_grid} and \eqref{e_LC} consisting of critical nodes  $i\in\hd{\mathcal{M}_c}$ and   ordinary nodes  $k\in\hd{\mathcal{M}_o}$. In steady state, voltage consensus of critical nodes can be achieved, i.e., $V_i=V_{rat}, \forall i\in\hd{\mathcal{M}_c}$, if and only if    
       $ \hd{\bm{I}_c}-\bm Y_{12}\bar{\bm Y}_{22}^{-1}\hd{\bm I_o}=V_{rat}\bar{\bm Y}_{22}|\bar{\bm Y}\mathbb 1_m,$
where $\hd{\bm{I}_c}=[I_1,\cdots,I_m]^T$, $\bar{\bm Y}_{22}|\bar{\bm Y}=\bar{\bm Y}_{11}-\bm Y_{12}\bar{\bm Y}_{22}^{-1}\bm Y_{21}$ is Schur complement of $\bar{\bm Y}_{22}$ with respect to  $\bar{\bm Y}$.
\end{lemma}
Lemma \ref{ft_conflict_2} establishes a  basis for achieving voltage consensus in critical nodes.
To enable voltage-current compromise, inspired by  \eqref{e_accurateI_consensus}, a trade-off factor is incorporated into the generalized consensus control protocol
\begin{equation}\label{e_Ipu_consensus}
  \dot{x}_i  =\sum^N_{j=1}a_{ij}\left(\frac{I_j}{I_{r_j}}-\frac{I_i}{I_{r_i}}\right), \forall i\in\mathcal{M},
\end{equation}
with
\begin{subequations}\label{e_newcompromise}
\begin{align}
  I_{r_k} & =\left\{
  \begin{aligned}
   &\theta I^*_k+\left(1-\theta\right)I_{b_k}, \quad \forall k\in\hd{\mathcal{M}_c}, \\
   &I^*_k, \quad\forall k\in\hd{\mathcal{M}_o},
\end{aligned}
\right.\label{e_newcompromise-b}\\
  \hd{\bm I_{b_c}}&=[I_{b_1},\cdots,I_{b_m}]^T = \omega V_{rat}\bar{\bm Y}_{22}|\bar{\bm Y}\mathbb 1_m+\bm Y_{12}{\bar{\bm Y}_{22}}^{-1}\hd{\bm I^*_o}, \label{e_newcompromise-a}
\end{align}
\end{subequations}
where $\theta\in[0,1]$ is a trade-off factor indicating the compromised degree; $\bm I^*_o=[I^*_{m+1},\cdots,I^*_{N}]^T$ is the rated current vector of  ordinary nodes; \hdba{$\omega$ is a design parameter for adjusting} the output currents of ordinary nodes. \hdba{The tuning rules of $\theta$ and $\omega$ are provided in Proposition \ref{prop_V_lim2}.}
\begin{remark}
Note that \eqref{e_Ipu_consensus} is generally a part of the control law $u_i$.
  For control schemes incorporating \eqref{e_accurateI_consensus} (e.g.,  \cite{NasirianFL2015-TPE,Tucci2018,nahata2020consensus,Liuzj2024TPS}),  replacing the constant $I^*_i$ in \eqref{e_accurateI_consensus} with the adjustable $I_{r_i}$ in \eqref{e_newcompromise-b} can upgrade the performance of the control schemes from accurate current sharing to a voltage-current compromise. Generally, this performance improvement does not affect the stability analysis of the closed-loop system, and thus is omitted.
\end{remark}
Let droop control serve as the primary control of the DC microgrid, receiving \hdba{the compensation} signal $x_i$ from the secondary control \eqref{e_Ipu_consensus}. Additionally, an integral signal $y_i$ is adopted to regulate the average voltage  to the rated value. Then  the distributed controller $u_i$ is proposed as
\begin{align}
    u_i &=V_{rat}-r_i I_i+x_i+y_i, \forall i\in\mathcal{M},\label{e_droop+distributed}\\
    \dot{y_i} &=V_{rat}-\tilde{V}_i,\label{e_integ}
\end{align}
where $r_i$ is the droop coefficient.

\subsection{Steady state analysis}
\begin{thm}\label{thm_cri_compromise}
Consider the DC microgrid \eqref{e_source}, \eqref{e_grid} and \eqref{e_LC} consisting of the critical nodes  $i\in\hd{\mathcal{M}_c}$ and the ordinary nodes  $k\in\hd{\mathcal{M}_o}$. These two classes of nodes are governed by \hdb{\eqref{e_cnode_obsv}, \eqref{e_cnode_obsv-b}, \eqref{e_Ipu_consensus}, \eqref{e_newcompromise}, \eqref{e_droop+distributed}, and \eqref{e_integ}}. Then, the following statements hold for steady state:
\begin{enumerate}[(a)]
  \hdb{\item\label{thm1_a} Average voltage regulation among critical nodes is achieved, i.e.,
  $\frac{1}{m}\sum^m_{j=1}V_j=V_{rat}.$}
  \item\label{thm1_b} For a given $\theta$, the output currents of DGs satisfy
       \begin{equation}\label{e_thm2_Iconsen}
          \frac{I_i}{I_{r_i}}=\frac{I_k}{I^*_k}=\frac{1}{\mu\theta+\omega (1-\theta)}, \forall i\in\hd{\mathcal{M}_c}, k\in\hd{\mathcal{M}_o}
        \end{equation}
        where $\mu=\frac{1}{mV_{rat}}\mathbb 1_m^T\left(\bar{\bm Y}_{22}|\bar{\bm Y}\right)^{-1}\left(\hd{\bm I^*_c}-\bm Y_{12}\bar{\bm Y}_{22}^{-1}\hd{\bm I^*_o}\right)$ with $\bm I^*_c=[I^*_1,\cdots,I^*_m]^T$ being the rated current vector of critical nodes.
  \item\label{thm1_c} Let $\bm\Delta^V_c(\theta)$ and $\bm\Delta^I_c(\theta)$ denote the
voltage and current deviation ratio of  critical nodes, respectively. Then, $\left|\bm\Delta^V_c(\theta)\right|$ and $\left|\hd{\bm \Delta^I_c}(\theta)\right|$ are monotonically increasing and decreasing, respectively, on $\theta\in[0,1]$, and
      \begin{subequations}\label{e_deltaV1}
      \begin{align}
     \hd{ \bm\Delta^V_c}(\theta) &= \frac{\theta}{(\mu-\omega)\theta+\omega}\bm\Psi_1 \label{e_deltaV1_1}\\
     \hd{ \bm\Delta^I_c}(\theta) &=\frac{1-\theta}{\theta\left({1}/\hd{{\bar I^{pu}_{b_c}}}-1\right)+1}\hd{\bm \Delta^I_{b_c}},\label{e_deltaV1_2}
      \end{align}
      \end{subequations}
      where $\bm\Psi_1=\frac{1}{V_{rat}}\left(\bar{\bm Y}_{22}|\bar{\bm Y}\right)^{-1} \left(\hd{\bm I^*_c}-\bm Y_{12}{\bar{\bm Y}_{22}}^{-1}\hd{\bm I^*_o}\right)-\mu \mathbb 1_m,$ $\hd{\bar I^{pu}_{b_c}}=\frac{1}{m}\mathbb 1_m^T\hd{\bm I^{pu}_{b_c}}$ and $\hd{\bm \Delta^I_{b_c}} =(\hd{\bm I^{pu}_{b_c}}-\hd{\bar I^{pu}_{b_c}}\mathbb 1_m)/\hd{\bar I^{pu}_{b_c}}$ with $\hd{\bm I^{pu}_{b_c}}=diag(\hd{\bm I^*_c})^{-1}\hd{\bm I_{b_c}}.$ Particularly, for $\theta\in[0,1]$, we have
\begin{equation}\label{e_cri_DeltaV_minmax}
  \qquad\;\left\{
    \begin{aligned}
      \min_{\theta}\left|\hd{\bm\Delta^V_c}(\theta)\right| &=\bm\Delta^V_c(0)=\mathbb 0\\
      \max_{\theta}\left|\hd{\bm\Delta^V_c}(\theta)\right| &=\left|\bm\Delta^V_c(1)\right|=\frac{1}{\mu}\left|\bm\Psi_1\right|,
    \end{aligned}
  \right.
\end{equation}
\begin{equation*}
  \left\{
    \begin{aligned}
      \min_{\theta}\left|\hd{\bm\Delta^I_c}(\theta)\right| &=\hd{\bm\Delta^I_c}(1)=\mathbb 0\\
      \max_{\theta}\left|\hd{\bm\Delta^I_c}(\theta)\right| &=\left|\hd{\bm\Delta^I_c}(0)\right|=\left|\hd{\bm\Delta^I_{b_c}}\right|.\;
    \end{aligned}
  \right.
\end{equation*}
\item\label{thm1_d} Voltage $\bm V_o=[V_{m+1},\dots, V_N]$ of ordinary nodes is
    \begin{equation*}
\hd{\bm V_o}=\frac{1}{\theta(\mu-\omega)+\omega}\left(\bm\Omega_2-\frac{\omega}{\mu-\omega}\bm\Omega_1\right)+\frac{1}{\mu-\omega}\bm\Omega_1,
\end{equation*}
where $\bm\Omega_1=-(\bar{\bm Y}_{11}|\bar{\bm Y})^{-1}\bm Y_{21}{\bar{\bm Y}_{11}}^{-1}(\hd{\bm I^*_c}-\hd{\bm I_{b_c}})$ and $\bm\Omega_2=(\bar{\bm Y}_{11}|\bar{\bm Y})^{-1}(\hd{\bm I^*_o}-\bm Y_{21}{\bar{\bm Y}_{11}}^{-1}\hd{\bm I_{b_c}})$. Each entry of $\hd{\bm V_o}$ is a monotone function on $\theta\in[0,1]$.
\end{enumerate}
\end{thm}
This result shows that the degree of voltage-current compromise of critical nodes can be adjusted by tuning $\theta$.
The monotonicity of $\left|\bm\Delta^V_c(\theta)\right|$ 
guarantees 
that 
strict voltage deviation at critical nodes can be achieved.

\begin{remark}
     Theorem \ref{thm_cri_compromise} implies that
accurate current sharing of ordinary nodes can always be achieved for any $\theta$, \hdba{while critical nodes can only achieve accurate current sharing when $\theta=1$; in addition, critical nodes achieve voltage consensus when $\theta=0$
 and a voltage-current compromise when $\theta\in(0,1)$}.
 \end{remark}
\subsection{Voltage deviation restriction for critical nodes}
The following proposition
provides guidelines for designing $\theta$ and $\omega$ based on a given $\Gamma_V$.
\begin{proposition}\label{prop_V_lim2}
Consider a DC microgrid with the same conditions as those in Theorem \ref{thm_cri_compromise}. The following statements hold.
\begin{enumerate}[(a)]
\item\label{prop2_a} Given a positive $\Gamma_V$, voltage deviation restriction for the critical nodes is achieved when $\theta\in\left[0,\theta_d\right]$, where $\theta_d$ is designed as
\begin{equation*}
  \theta_d =\left\{
    \begin{aligned}
      &\qquad 1, \qquad when\quad \left\|\hd{\bm\Delta^V_c}(1)\right\|_{\infty}<\Gamma_V\\
      &\frac{\omega\Gamma_V/(\mu-\omega)}{\|\bm\Psi_1\|_\infty-\Gamma_V},\, when\; 0<\Gamma_V\le\left\|\hd{\bm\Delta^V_c}(1)\right\|_{\infty}\\
      &\qquad 0, \qquad when\quad \Gamma_V=0.
    \end{aligned}
  \right.
\end{equation*}
\item\label{prop2_b}  Given $\theta\in[0,1)$, the adjustable range of $\omega$ is $$\mathcal{W}=\left[\max_i\left\{\frac{\zeta_i}{\nu_i}\right\},+\infty\right), i\in\hd{\mathcal{M}_c}, \forall \nu_i>0,$$
where $[\zeta_1, \cdots, \zeta_m]^T=\frac{-\theta}{1-\theta}\hd{\bm I^*_c}-\bm Y_{12}{\bar{\bm Y}_{22}}^{-1}\hd{\bm I^*_o}$, and $[\nu_1,\cdots, \nu_m]^T=V_{rat}\bar{\bm Y}_{22}|\bar{\bm Y}\mathbb 1_m$.
  \item\label{prop2_c}  The output per-unit currents of DGs at the ordinary nodes $\hd{\bm I^{pu}_o}$ are monotonically decreasing on $\omega\in\mathcal{W}$.
  \item\label{prop2_d}  The sum of the output currents of DGs at the critical nodes $\mathbb{1}^T_m\hd{\bm I_c}$ is monotonically increasing on $\omega\in\mathcal{W}$ if $\theta\mathbb{1}_m^T\hd{\bm I^*_c}+\left(1-\theta\right)\mathbb{1}_m^T\bm Y_{12}\bar{\bm Y}_{22}^{-1}\hd{\bm I^*_o}<0$.
\end{enumerate}
\end{proposition}

\begin{remark}
According to Theorem \ref{thm_cri_compromise} and Propositions \ref{prop_Vcri_obser} and \ref{prop_V_lim2}, Problem \ref{plm_1} can be addressed by \hdb{the control scheme \eqref{e_cnode_obsv}, \eqref{e_cnode_obsv-b}, \eqref{e_Ipu_consensus}, \eqref{e_newcompromise} and \eqref{e_droop+distributed}, \eqref{e_integ}.} \hwz{Specifically, \eqref{e_thm2_Iconsen} in Theorem \ref{thm_cri_compromise} and Statement (\ref{prop2_b}) in Proposition \ref{prop_V_lim2} show that Objective (\ref{obj-c}) is achieved; Statement (\ref{thm1_a}) in Theorem \ref{thm_cri_compromise} and Statement (\ref{prop2_a}) in Proposition \ref{prop_V_lim2} jointly ensure Objective (\ref{obj-a}); Statement (\ref{thm1_c}) in Theorem \ref{thm_cri_compromise} illustrates the achievement of Objective (\ref{obj-b}).}
\end{remark}

\section{Simulation examples}\label{sec_simulation}
\begin{figure}[htbp]
  \centering
  \includegraphics[width=6cm]{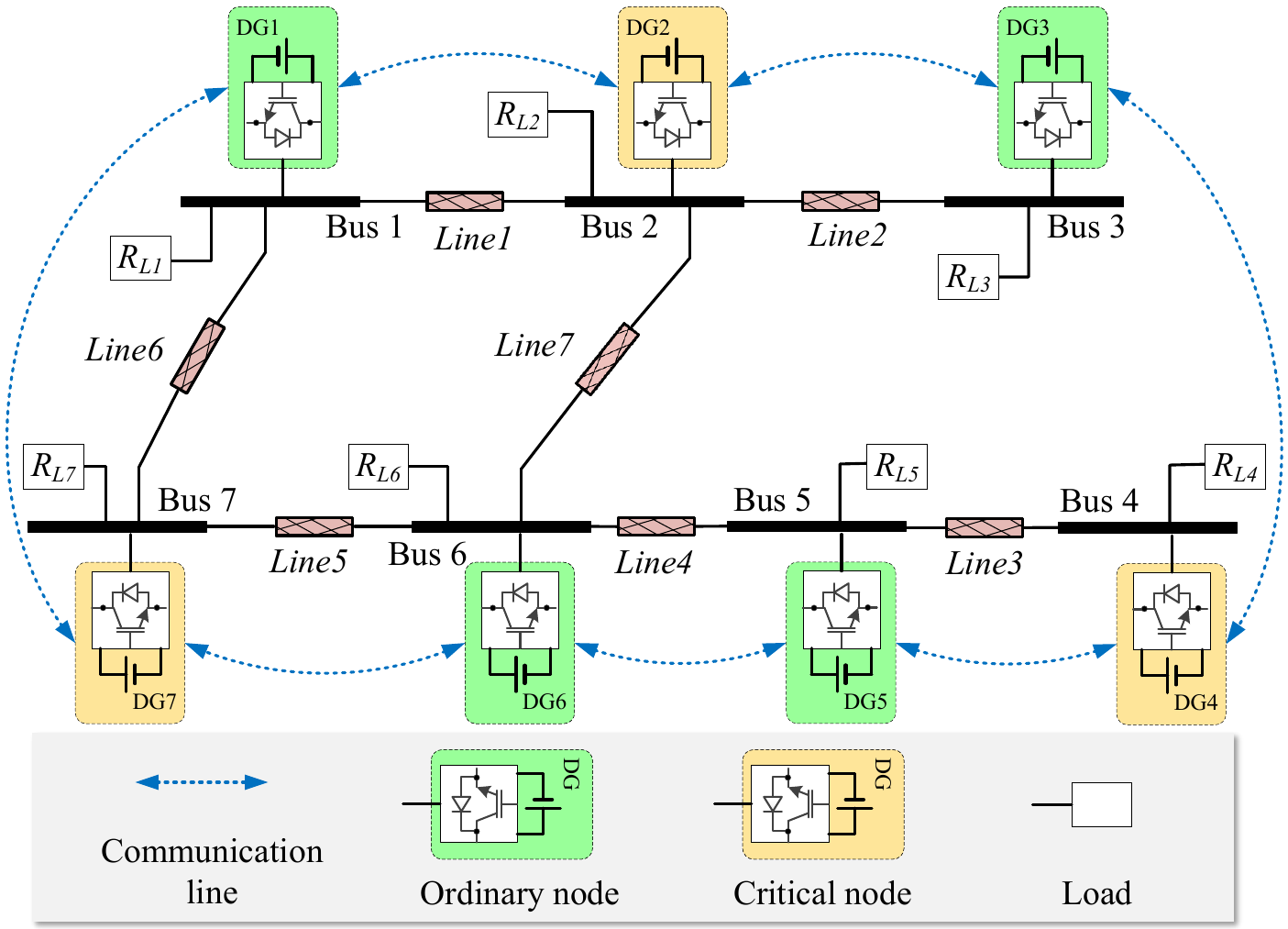}\\
  \caption{DC microgrid with critical nodes \{2,4,7\} and ordnary nodes \{1,3,5,6\}} \label{7bus_MG}
\end{figure}

Consider a DC microgrid composed of $7$ DGs, as shown in Fig. \ref{7bus_MG}.
Without loss of generality, let the weight of each communication link be $a_{ij}=20$, where a greater $a_{ij}$ is beneficial to faster convergence without changing the steady-state value \citep{FLewis2014Book-distributed}.

The rated voltage of the microgrid is set to ${V_{rat}=380\mathrm{V}}$.
Let $\bm I^*=diag(30, 30, 20, 20, 40, 40, 40)\ \mathrm{A},$
  $ \bm L_t =diag(2, 2.2, 1.8, 2.5, 3, 2.6, 2.3)\ \mathrm{mH},$ $
   \bm C_t =diag(3, 2.5, \text{2.8},\\ 2.5, 2.3, 3, 2.6)\ \mathrm{mF},$ $ 
   \bm R_L =diag(50, 20, 26, 35, 38, 23, 40)\ \Omega.$
The parameters of power lines are $R_{l_1}=2\Omega$, $L_{l_1}=20\mathrm{\mu H}$; $R_{l_2}=2.4\Omega$, $L_{l_2}=25\mathrm{\mu H}$; $R_{l_3}=2\Omega$, $L_{l_3}=20\mathrm{\mu H}$; $R_{l_4}=4\Omega$, $L_{l_4}=35\mathrm{\mu H}$; $R_{l_5}=2\Omega$, $L_{l_5}=20\mathrm{\mu H}$; $R_{l_6}=2\Omega$, $L_{l_6}=20\mathrm{\mu H}$; $R_{l_7}=2\Omega$, $L_{l_7}=20\mathrm{\mu H}$.

The simulation examples consist of two cases, where the control laws \hdb{\eqref{e_cnode_obsv}, \eqref{e_cnode_obsv-b}, \eqref{e_Ipu_consensus}, \eqref{e_newcompromise}, \eqref{e_droop+distributed}, and \eqref{e_integ}} are used.

\emph{Case I: \bf{Compromised control for critical nodes}}


For critical nodes, the objective is to restrict the voltage deviation within $[0, \Gamma_V]$, where $\Gamma_V=2\%$, meanwhile maintaining the current sharing performance of ordinary nodes. The following results show how $\theta$ and $\omega$ affect the system performance.

\begin{figure}[htbp]
  \centering
  \includegraphics[width=8.4cm]{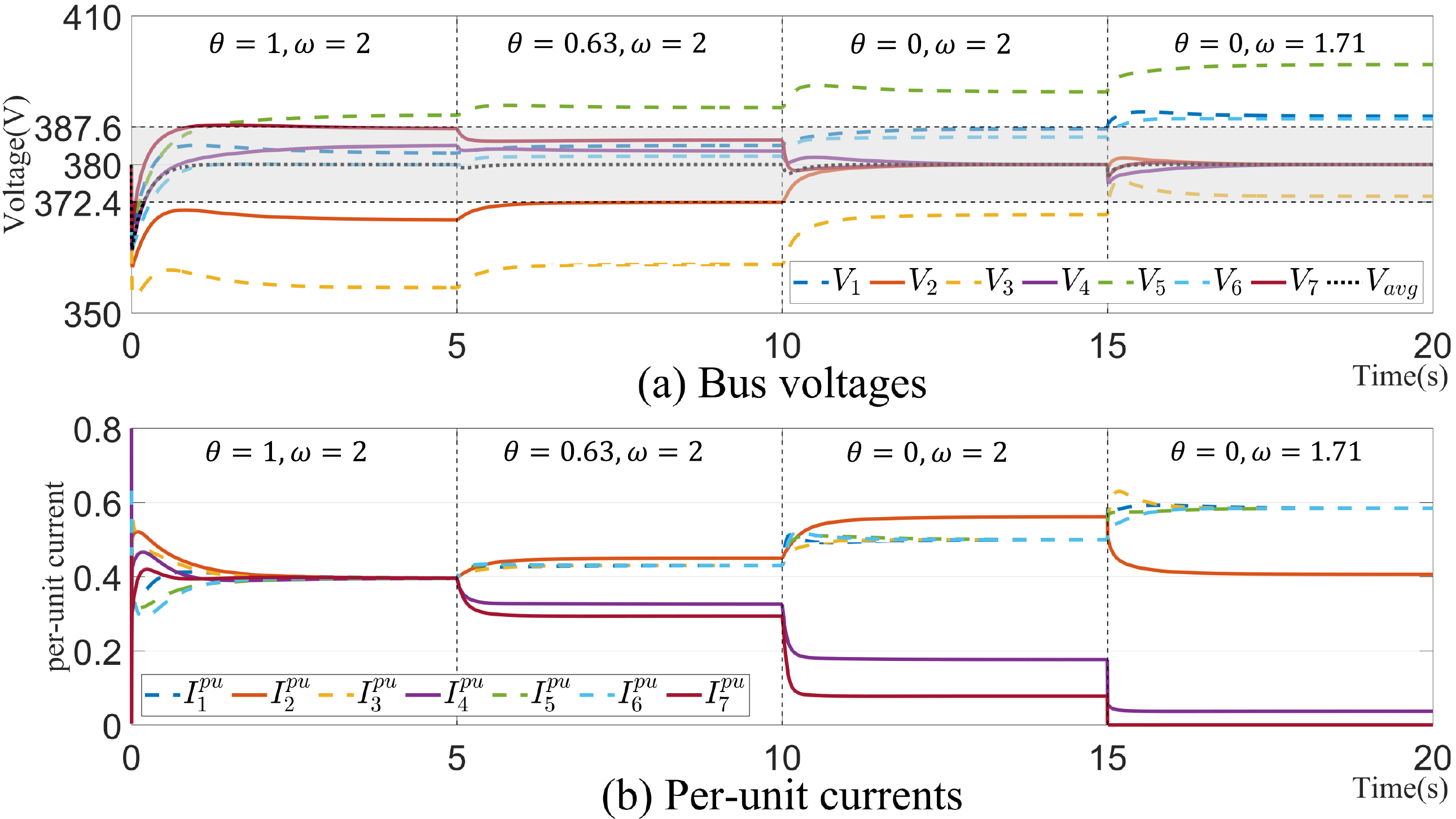}\\
  \caption{Simulation results of Case I. The shaded area corresponds to the admissible voltage range for critical nodes.} \label{fig:cri_comprised}
\end{figure}

During $t\in[0,5)s$, let $\theta=1$, $\omega=2$ in controller \eqref{e_newcompromise}. In steady state, Fig. \ref{fig:cri_comprised}(b) shows current sharing for all nodes is achieved, while Fig. \ref{fig:cri_comprised}(a) shows the voltage deviation of node 7 is beyond the admissible range (shaded area).

During $t\in[5,10)s$,  we set $\theta=0.63$ and $\omega=2$. In fact, according to Theorem \ref{thm_cri_compromise} and Proposition \ref{prop_V_lim2}, the $\Delta^V_{max}$ of critical nodes can be restricted within $[0, 0.02]$ by setting $\theta\in [0,0.63]$ when $\omega=2$, which is shown in Fig. \ref{fig:cri_comprised}(a). At the same time, the current sharing performance of critical nodes is degraded, which shows a compromise between voltage consensus and current sharing of critical nodes. It is noted from Fig. \ref{fig:cri_comprised}(b) that the current sharing of ordinary nodes is still maintained.

During $t\in [10,15)s$, we set $\theta=0$ and $\omega=2$. Voltage consensus of all critical nodes is achieved and $\lim_{t \to \infty}V_i(t)=V_{rat}, \forall i\in \hd{\mathcal{M}_c}$ (see Fig. \ref{fig:cri_comprised}(a)). In this case, the critical nodes have the worst current sharing performance (see Fig. \ref{fig:cri_comprised}(b)). At the same time, the current sharing is still achieved for ordinary nodes.

During $t\in [15,20]s$, we keep $\theta=0$ and tune $\omega$ from 2 to 1.71. It is observed that the output currents of ordinary nodes increase (see Fig. \ref{fig:cri_comprised}(b)). Meanwhile, the output currents of critical nodes decrease accordingly. Also note that when $\omega = 1.71$, $I^{pu}_7 = 0$, the per-unit currents of ordinary nodes reach the admissible maximum $I^{pu}_i = 0.585, i = \{1, 3, 5, 6\}$ (see Fig. \ref{fig:cri_comprised}(b)). In this phase, since $\theta=0$, the voltage consensus for critical nodes still holds, i.e., $\lim_{t\to \infty}V_i(t)=V_{rat}, \forall i\in \hd{\mathcal{M}_c}$ (see Fig. \ref{fig:cri_comprised}(a)).

Moreover, we can see that the average voltage of critical nodes can always be regulated to the rated value of 380V (see Fig. \ref{fig:cri_comprised}(a)), and current sharing for ordinary nodes can also be achieved (see Fig. \ref{fig:cri_comprised}(b)), both independent of the choices of $\theta$ and $\omega$.

\emph{Case II: \bf{\hwzh{Plug-and-play (PnP)} operation  of DGs}}

The configurations of critical nodes and ordinary nodes are the same as in Case I.
Let $\theta=0$ and $\omega=2$. This setting corresponds to accurate voltage consensus of critical nodes and accurate current sharing of ordinary nodes.
To illustrate the performance of the proposed control laws under the PnP of DGs connected to ordinary node and critical node respectively, we perform two simulations.

\emph{(a) \bf{PnP of DG at ordinary node}}

\begin{figure}[htbp]
  \centering
  \includegraphics[width=8.4cm]{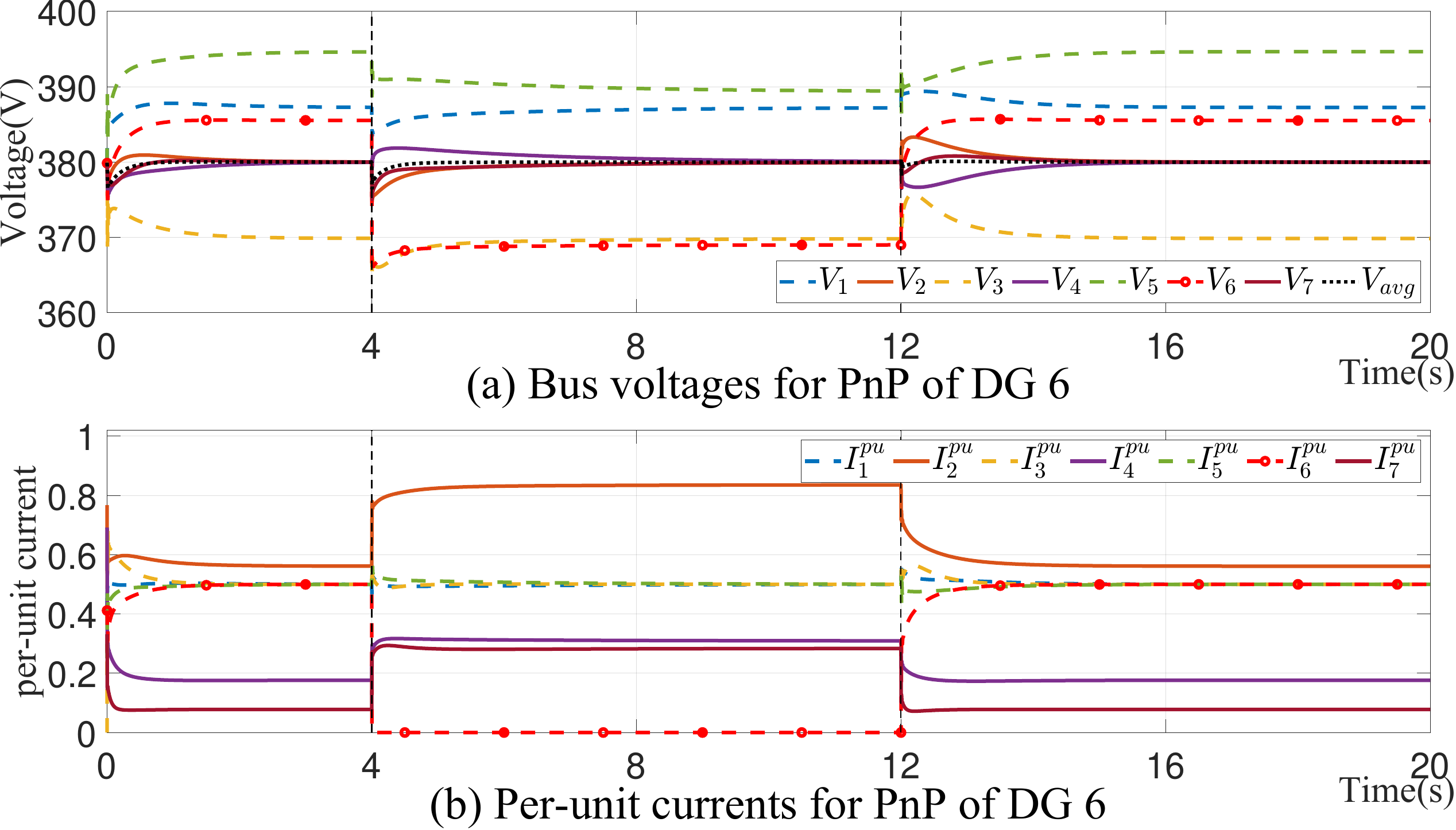}\\
  \caption{Simulation results for Case II (a): \ref{fig:ordinary_pnp}(a) and \ref{fig:ordinary_pnp}(b) depict the PnP process of node 6, an ordinary node. At $t=4s$, the DG of nodes 6 is unplugged, and at $t=12s$, it is plugged back into the microgrid.} \label{fig:ordinary_pnp}
\end{figure}

During $t\in[0,4)s$, all nodes are normally operated.
Figure  \ref{fig:ordinary_pnp}(a) shows that all critical node voltages are regulated to 380V, i.e., accurate voltage consensus is achieved, while currents are accurately shared among all ordinary nodes (i.e., 1, 3, 5, 6) (see Fig. \ref{fig:ordinary_pnp}(b)).

During $t\in[4,12)s$, the DG of node 6 is plugged out from the DC microgrid. Then the output current of DG 6 becomes zero immediately at $t=4s$ (see Fig. \ref{fig:ordinary_pnp}(b)), and meanwhile, the bus voltage of node 6 is dropped from 385.5V to 369V (see Fig. \ref{fig:ordinary_pnp}(a)). However, voltage consensus of critical nodes and accurate current sharing of the rest ordinary nodes are still achieved at steady state.

During $t\in[12,20]s$, the DG of node 6 is plugged back into the DC microgrid.
It can be observed from Fig. \ref{fig:ordinary_pnp} that bus voltages and per-unit currents of all nodes are successfully restored.

\emph{(b) \bf{PnP of DG at critical node}}

\begin{figure}[htbp]
  \centering
  \includegraphics[width=8.4cm]{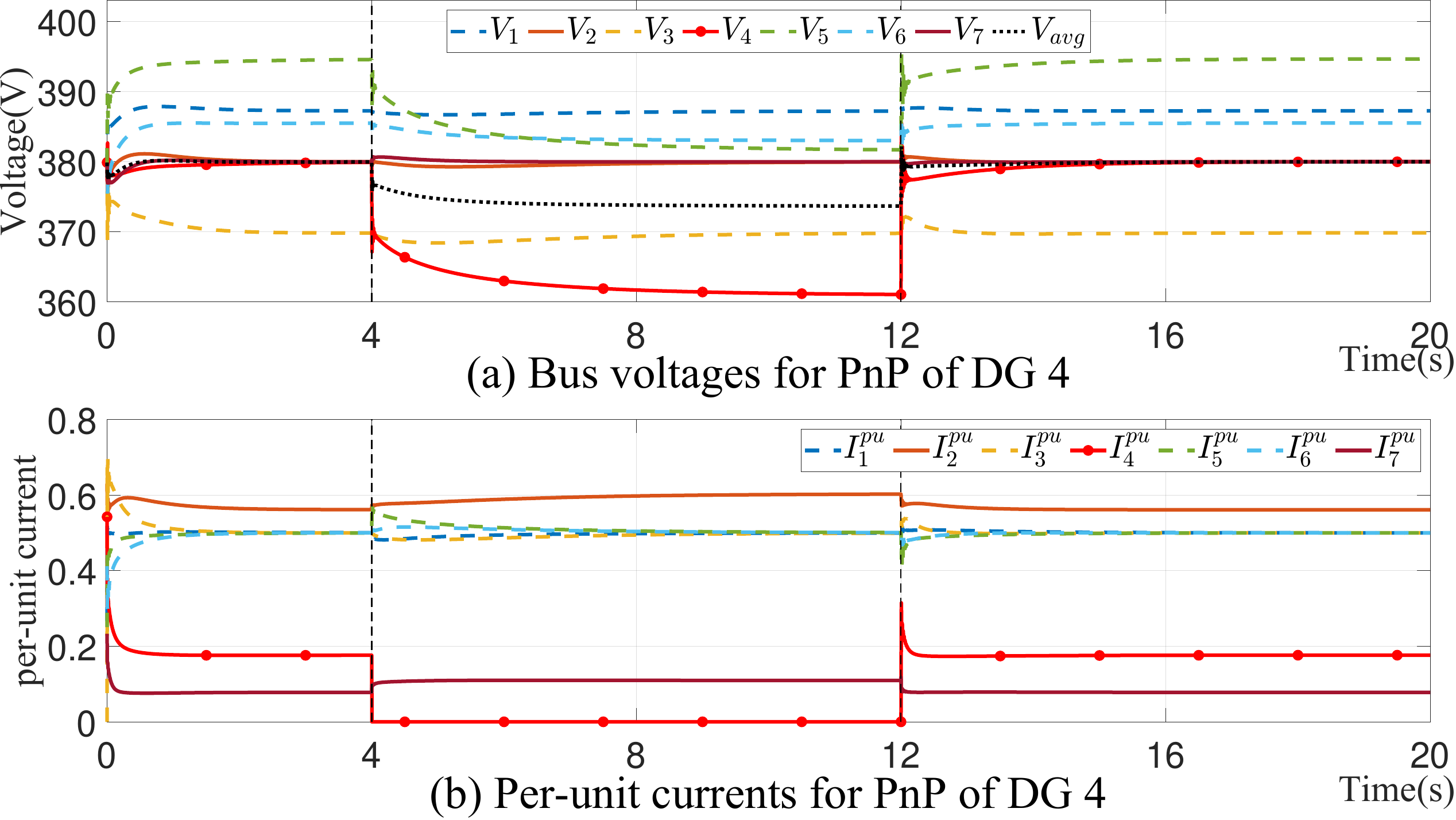}\\
  \caption{Simulation results for Case II (b): \ref{fig:critical_pnp}(a) and \ref{fig:critical_pnp}(b) depict the PnP process of node 4, a critical node. At $t=4s$, the DG of nodes 4 is unplugged, and at $t=12s$, it is plugged back into the microgrid.} \label{fig:critical_pnp}
\end{figure}

To unplug the DG \hdba{at} critical node 4, the critical node must first be degraded as an ordinary node, and all other critical node controllers should be updated accordingly. Then, the DG of this node can be unplugged.  Reconnecting follows the exact reverse procedure. Simulation results are shown in Fig. \ref{fig:critical_pnp}(a) and \ref{fig:critical_pnp}(b). \hwzh{During $t\in [4, 12]s$, the output
current of DG 4 drops to zero immediately at $t=4s$, and meanwhile, the bus
voltage of node 4 is dropped continuously from 380V to 362V. However, voltage consensus of the
rest of the critical nodes \{2, 7\} and accurate current sharing of the ordinary nodes can still be
achieved at the steady state. After $t=12s$, both voltages and per-unit currents recover. }
\section{Conclusion}\label{sec_conclusion}
This paper proposed a voltage-current compromised control scheme for multi-bus DC microgrids. By introducing voltage deviation ratio and current deviation ratio, we formulated three objectives of accurate current sharing, voltage consensus and voltage-current compromise in a unified manner, and figured out a necessary and sufficient condition for the conflict between accurate current sharing and voltage consensus. For a class of DC microgrids \hwzh{consisting of} critical nodes and ordinary nodes, we proposed a distributed compromised control law such that the bus voltage deviation of all critical nodes can be arbitrarily restricted without sacrificing accurate current sharing of those ordinary nodes, and the degree of compromise between voltage deviation and current sharing can be easily adjusted by a single trade-off parameter. It is worth noting that the proposed algorithm can extend various existing current sharing control methods to a compromised objective.
This control algorithm also works under PnP settings.
\section*{Appendix}
\appendix
\section{Proof of Fact \ref{ft_conflict}}\label{apd_ft1}
Letting $\dot{\bm I_l}=\mathbb 0$, $\dot{\bm I}=\mathbb 0$ and $\dot{\bm V}=\mathbb 0$ in \eqref{e_grid} and \eqref{e_LC} and eliminating $\bm I_l$ yields the steady state model 
\begin{subequations}\label{e_ssmdl}
\begin{align}
   \bm V_t-\bm V  \label{e_ssmdl-a} &=\mathbb 0\\
   \bm I-\bm Y_L\bm V-\bm Y\bm V &=\mathbb 0, \label{e_ssmdl-b}
\end{align}
\end{subequations}
where $\bm Y=\bm B\bm R_l^{-1}\bm B^T$ is the admittance matrix of the DC microgrid, and
it is an irreducible Laplacian with nonnegative eigenvalues \citep{dorfler2018ProcIEEE-electr}. 
Then  $\left(\bm Y+\bm Y_L\right)^{-1}>\mathbb{0}$ if and only if $\bm Y_L\ge \mathbb 0$ and $\bm Y_L\ne \mathbb 0$. \hwz{Let $\Delta \bm{V} = \bm V - V_{rat}\mathbb 1_N$}.
Then \eqref{e_ssmdl-b} can be put as
$\bm I =(\bm Y+\bm Y_L)(V_{rat}\mathbb{1}_N+\Delta \bm{V})$.
Moreover, by noticing  $\bm{\Delta}^V=\Delta\bm V/V_{rat}$, we have
\begin{equation}\label{e_deltaV_deltaI}
  \bm{\Delta}^V=\left(\bm Y+\bm Y_L\right)^{-1}\left(\frac{\bm I}{V_{rat}}-\bm Y_L\mathbb 1_N\right).
\end{equation}
\hd{It implies that $\bm \Delta^V= \mathbb 0$ if and only if $\bm I=V_{rat}\bm Y_L\mathbb 1_N$.}
Moreover, when accurate current sharing 
is attained, 
we have $\bm I=\alpha\bm I^*$ with $\alpha$ being a positive constant.
Thus $\bm \Delta^V=\mathbb 0$ and $\bm \Delta^I=\mathbb 0$ if and only if $\bm I^*=\frac{V_{rat}}{\alpha}\bm Y_L\mathbb 1_N$.

\section{Proof of Proposition \ref{prop_Vcri_obser}}\label{apd_prop_Vcri_obser}
One can rewrite \eqref{e_cnode_obsv} and \eqref{e_cnode_obsv-b} in compact form as follows,
  $\left[
     \begin{array}{c}
       \dot{\tilde{\bm V}}_1 \\
       \mathbb 0 \\
     \end{array}
   \right]=\left[\begin{array}{c}
                     \dot{\bm V}_1 \\
                     \mathbb 0 \\
                 \end{array}
           \right]-\left[
              \begin{array}{cc}
                \mathcal L_{11} & \mathcal L_{12} \\
                \mathcal L_{21} & \mathcal L_{22} \\
              \end{array}
            \right]\left[
                     \begin{array}{c}
                       \tilde{\bm V}_1 \\
                       \tilde{\bm V}_2 \\
                     \end{array}
                   \right].$
Then, it holds that
  $\dot{\tilde{\bm V}}_1=\dot{\bm V}_1-\mathcal L_{22}|\mathcal L\tilde{\bm V}_1.$
Since the graph associated with $\mathcal L$ is undirected and connected, according to \cite{dorfler2013TCSI}, $\mathcal L_{22}|\mathcal L$ is a symmetric Laplacian, and the graph associated with $\mathcal L_{22}|\mathcal L$ is undirected and connected.
Hence, according to \cite{Freeman2019MCS}, we have
$\lim_{t\to\infty}\tilde{V}_i=\frac{1}{m}\sum^m_{j=1}V_j, \hdbai{\forall} i\in\mathcal{M}.$

\section{Proof of Lemma \ref{ft_conflict_2}}
The steady state equation \eqref{e_ssmdl-b} can be rewritten as,
\begin{equation}\label{e_steady_grid}
   \left[\begin{array}{c}
               \hd{\bm I_c} \\
               \hd{\bm I_o} \\
            \end{array}
      \right]=   \left[\begin{array}{cc}
               \bar{\bm Y}_{11} &\bm Y_{12} \\
              \bm Y_{21} & \bar{\bm Y}_{22} \\
            \end{array}
      \right]\left[\begin{array}{c}
               \hd{\bm V_c} \\
               \hd{\bm V_o} \\
            \end{array}
      \right],
\end{equation}
where $\hd{\bm I_c,\bm V_c}\in\mathbb R^m$  are the steady state currents and voltages of critical nodes, respectively; and $\bm I_o,\bm V_o\in\mathbb R^{N-m}$ are similarly defined for ordinary nodes.
\hwz{A straightforward manipulation to \eqref{e_steady_grid} yields}
\begin{equation}\label{e_crinode_I1}
    \bm I_c=\bar{\bm Y}_{22}|\bar{\bm Y}\bm V_c+\bm Y_{12}\bar{\bm Y}_{22}^{-1}\bm I_o.
\end{equation}

Let \hwz{$\bm{\Delta}^{V_c}=\frac{\bm V_c}{V_{rat}}-\mathbb 1_m$}. Then \eqref{e_crinode_I1} can be put as
    $\bm I_c=V_{rat}\bar{\bm Y}_{22}|\bar{\bm Y}\mathbb{1}_m+V_{rat}\bar{\bm Y}_{22}|\bar{\bm{Y}}\bm{\Delta}^{V_c}+\bm Y_{12}\bar{\bm Y}_{22}^{-1}\bm I_o.$
Further,
$\bm{\Delta}^{V_c}=\frac{1}{V_{rat}}(\bar{\bm Y}_{22}|\bar{\bm{Y}})^{-1}\left(\bm I_c-V_{rat}\bar{\bm Y}_{22}|\bar{\bm{Y}}\mathbb 1_m-\bm Y_{12}\bar{\bm Y}_{22}^{-1}\bm I_o\right).$
\hd{Since $\bm Y^{-1}>\mathbb{0}$, it holds that $\left(\bar{\bm Y}_{22}|\bar{\bm Y}\right)^{-1}>\mathbb{0}$.}
Then it is clear that $\bm{\Delta}^{V_c}=\mathbb 0$ if and only if $\bm{I}_c-\bm Y_{12}\bar{\bm Y}_{22}^{-1}\bm I_o=V_{rat}\bar{\bm Y}_{22}|\bar{\bm Y}\mathbb 1_m$.

\section{Proof of Theorem \ref{thm_cri_compromise}}
\hdb{($a$)
According to Proposition \ref{prop_Vcri_obser}, it holds that $\tilde{V}_i=\frac{1}{m}\sum^m_{j=1}V_j$ in steady state. Considering $\dot{y}_i=0$ in \eqref{e_integ} \hwzh{(please refer to \cite{NasirianFL2015-TPE} for the detailed proof)}, we have
$\tilde V_i=\frac{1}{m}\sum^m_{j=1}V_j=V_{rat}.$}

($b$)
 Put \eqref{e_Ipu_consensus} into compact form as $\dot{\bm X}=-\mathcal{L}diag(\bm I_r)^{-1}\bm I$. Since $\mathcal{G(\mathcal{A})}$ is undirected and connected, $\mathcal{L} \mathbb{1}_N=0$ \citep{FLewis2014Book-distributed}. Therefore, since $\dot{\bm X}=0$ in steady state, $diag(\bm I_r)^{-1}\bm I=\alpha\mathbb{1}_N$.
It is straightforward to yield
$\frac{I_1}{I_{r_1}}=\cdots=\frac{I_m}{I_{r_m}}=\frac{I_{m+1}}{I^*_{m+1}}=\cdots=\frac{I_N}{I^*_N}=\alpha.$
Substituting $\hd{\bm I_c}=\alpha \hd{\bm I_{r_c}}=\alpha (\theta \hd{\bm I^*_c}+(1-\theta)\hd{\bm I_{b_c}})$ and $\hd{\bm I_o}=\alpha \hd{\bm I^*_o}$ into \eqref{e_crinode_I1} yields
$\alpha \left(\theta (\hd{\bm I^*_c}-\hd{\bm I_{b_c}})+\hd{\bm I_{b_c}}\right)=\bar{\bm Y}_{22}|\bar{\bm Y}\hd{\bm V_c}+\bm Y_{12}{\bar{\bm Y}_{22}}^{-1}\alpha \hd{\bm I^*_o}.$
In addition, voltage balancing for critical nodes implies $\hd{\bm V_c}=V_{rat}(\mathbb 1_m+\hd{\bm{\Delta}^V_c})$ with $\mathbb 1_m^T\hd{\bm{\Delta}^V_c}=0$. Consequently, 
\begin{equation}\label{e_alpha_cal}
\begin{aligned}
\begin{split}
  \alpha \left(\bar{\bm Y}_{22}|\bar{\bm Y}\right)^{-1}\left(\theta(\hd{\bm I^*_c}-\hd{\bm I_{b_c}})+\omega V_{rat}\bar{\bm Y}_{22}|\bar{\bm Y}\mathbb 1_m\right) \\
  =V_{rat}\left(\mathbb 1_m+\hd{\bm{\Delta}^V_c}\right).
\end{split}
\end{aligned}
\end{equation}
Pre-multiplying both sides of \eqref{e_alpha_cal} by $\mathbb 1_m^T$ gives
  $\alpha=\frac{1}{\theta\mu+\omega(1-\theta)}.$
Since $\omega>0, \theta\in[0,1]$ and $\mu>0$, it is obvious that $\alpha>0$.

($c$)
Substituting $\alpha$ in \eqref{e_alpha_cal} and performing straightforward manipulation shows that \eqref{e_deltaV1_1} holds.
Since $\alpha> 0$, it holds that $\frac{\theta}{\mu\theta+\omega-\omega\theta}\ge 0$. The function $\left|\bm \Delta^V(\theta)\right|$ is monotonically increasing for $\theta\in[0,1]$. Furthermore, it is straightforward to obtain \eqref{e_cri_DeltaV_minmax}.

In steady state, $\hd{\bm I_c}=\alpha\left(\theta \hd{\bm I^*_c}+(1-\theta)\hd{\bm I_{b_c}}\right)$. Therefore, we have
$\hd{\bm I^{pu}_c}=\alpha\left(\theta\mathbb 1_N+(1-\theta)\hd{\bm I^{pu}_{b_c}}\right),$
where $\hd{\bm I^{pu}_{b_c}}=diag(\hd{\bm I^*_c})^{-1}\hd{\bm I_{b_c}}$.
By Definition 1, we have $\hd{\bm I^{pu}_c}=\hd{I^{pu}_{{avg}_c}}(\hd{\bm\Delta^{I_c}}+\mathbb 1_m)$ with $\mathbb 1_m^T\hd{\bm\Delta^{I_c}}=0$.
Consequently,
\begin{equation}\label{e_DeltaI_cal}
  \hd{I^{pu}_{{avg}_c}}(\hd{\bm\Delta^{I_c}}+\mathbb 1_m)=\alpha\left(\theta\mathbb 1_m+(1-\theta)\hd{\bm I^{pu}_{b_c}}\right),
\end{equation}
which further implies, by left multiplying \eqref{e_DeltaI_cal} by $\mathbb 1_m^T$, that
$I^{pu}_{{avg}_c}=\alpha\left(\theta\left(1-\hd{\bar I^{pu}_{b_c}}\right)+\hd{\bar I^{pu}_{b_c}}\right),$
where $\hd{\bar I^{pu}_{b_c}}=\frac{1}{m}\mathbb 1_m^T\hd{\bm I^{pu}_{b_c}}$.
Hence, substituting $I^{pu}_{{avg}_c}$ into \eqref{e_DeltaI_cal} yields
\begin{equation}
  \hd{\bm \Delta^{I_c}}(\theta) 
  =\left( \frac{1}{\theta(1-\hd{\bar I^{pu}_{b_c}})+\hd{\bar I^{pu}_{b_c}}}-1\right)\frac{\hd{\bar I^{pu}_{b_c}}}{1-\hd{\bar I^{pu}_{b_c}}}\hd{\bm \Delta^{I_c}_b},
\end{equation}
where $\hd{\bm \Delta^{I_c}_b} =(\hd{\bm I^{pu}_{b_c}}-\hd{\bar I^{pu}_{b_c}}\mathbb 1_m)/\hd{\bar I^{pu}_{b_c}}$.
It is straightforward to conclude that $\hd{\bm\Delta^{I_c}}(\theta)$ is monotonous on $\theta\in[0,1]$.
Moreover, letting $\theta=1$ yields $\hd{\bm\Delta^{I_c}}(1)=\mathbb 0$. Thus, $|\hd{\bm\Delta^{I_c}}(\theta)|$ is monotonically decreasing on $\theta\in[0,1]$.

($d$)
According to \eqref{e_steady_grid},
$\bm I_o =\left(\bar{\bm Y}_{22}-\bm Y_{21}{\bar{\bm Y}_{11}}^{-1}\bm Y_{12}\right)\hd{\bm V_o}+\bm Y_{21}{\bar{\bm Y}_{11}}^{-1}\hd{\bm I_c}.$
Considering, in steady state, $\hd{\bm I_o}=\alpha\hd{\bm I^*_o}$ and $\hd{\bm I_c}=\alpha\left(\theta \hd{\bm I^*_c}+(1-\theta)\hd{\bm I_{b_c}}\right)$, we have
\begin{equation*}
\alpha\left( \hd{\bm I^*_o}-\bm Y_{21}{\bar{\bm Y}_{11}}^{-1}\left(\theta \hd{\bm I^*_c}+(1-\theta)\hd{\bm I_{b_c}}\right) \right)=\bar{\bm Y}_{11}|\bar{\bm Y}\hd{\bm V_o}.
\end{equation*}
Then, substituting $\alpha$ into the above equation yields
$\hd{\bm V_o}=\frac{1}{\theta\mu+\omega(1-\theta)}\left(\theta\bm\Omega_1+\bm\Omega_2\right).$
Furthermore, it is derived that each entry of $\hd{\bm V_o}$ is a monotone function for $\theta\in[0,1]$, and
$\hd{\bm V_o}=\frac{1}{\theta(\mu-\omega)+\omega}\left(\bm\Omega_2-\frac{\omega}{\mu-\omega}\bm\Omega_1\right)+\frac{1}{\mu-\omega}\bm\Omega_1.$
\section{Proof of Proposition \ref{prop_V_lim2}}\label{apd_prop_onode_current}
$(a)$ According to \eqref{e_cri_DeltaV_minmax}, it is trivial to obtain that $\theta_d=1$ (or $\theta_d=0$) if $\left\|\hd{\bm\Delta^{V_c}}(1)\right\|_{\infty}<\Gamma_V$ (or $\Gamma_V=0$).
Then, according to \eqref{e_deltaV1_1} and considering $\mu\theta_d+\omega(1-\theta_d)> 0$, for given $\Gamma_V$, if $\Gamma_V\in\left(0,\left\|\hd{\bm\Delta^{V_c}}(1)\right\|_{\infty}\right]$, it holds that
 $ \Gamma_V=\max_\theta\left\|\hd{\bm \Delta^{V_c}}(\theta)\right\|_\infty=\frac{\theta_d}{\left(\mu-\omega\right)\theta_d+\omega}\|\bm\Psi_1\|_\infty,$
which further implies
  $\theta_d=\frac{\omega\Gamma_V}{\|\bm\Psi_1\|_\infty-\Gamma_V(\mu-\omega)}.$

$(b)$ The range of $\omega$ is determined by the admissible range of the output currents of DGs.
According to the statement ($a$) of Theorem \ref{thm_cri_compromise}, in steady state, the output per-unit currents of DGs at critical nodes are
  $\bm I^{pu}_c 
  =\frac{\theta}{\mu \theta+\omega(1-\theta)}\mathbb 1_m+\frac{1-\theta}{\mu \theta+\omega(1-\theta)}diag(\bm I^*_c)^{-1}\bm I_{b_c}.$
Since $I^{pu}_i=\frac{I_i}{I^*_i}\ge 0, \forall i\in \mathcal{M}$, and $\hd{\bm I_{b_c}}=\omega V_{rat}\bar{\bm Y}_{22}|\bar{\bm Y}\mathbb 1_m+\bm Y_{12}{\bar{\bm Y}_{22}}^{-1}\hd{\bm I^*_o}$, we have
   $ \left(1-\theta\right)\left(\omega V_{rat}\bar{\bm Y}_{22}|\bar{\bm Y}\mathbb 1_m+\bm Y_{12}{\bar{\bm Y}_{22}}^{-1}\hd{\bm I^*_o}\right) \ge -\theta\hd{\bm I^*_c} .$
Further considering $\bar{\bm Y}_{22}|\bar{\bm Y}\mathbb 1_m\ge \mathbb 0$ and $\bar{\bm Y}_{22}|\bar{\bm Y}\mathbb 1_m\ne \mathbb 0$ yields
   $ \omega \in \left[\max_i\left\{\frac{\zeta_i}{\nu_i}\right\},+\infty\right), i\in\hd{\mathcal{M}_c}, \forall \nu_i\ne 0.$
where $[\zeta_1,\cdots,\zeta_m]^T=\frac{-\theta}{1-\theta}\hd{\bm I^*_c}-\bm Y_{12}{\bar{\bm Y}_{22}}^{-1}\hd{\bm I^*_o}, [\nu_1,\cdots,\nu_m]^T=V_{rat}\bar{\bm Y}_{22}|\bar{\bm Y}\mathbb 1_m$.

$(c)$ Since $\hd{\bm I^{pu}_o}=\frac{1}{\mu \theta+\omega(1-\theta)}$, it is straightforward to show the monotonic decreasing property of $\hd{\bm I^{pu}_o}$ on $\omega\in\mathcal{W}$.

$(d)$\\
  $\mathbb{1}^T_m\hd{\bm I_c}  =\frac{\theta}{\mu \theta+\omega(1-\theta)}\mathbb{1}_m\hd{\bm I^*_c}+\frac{1-\theta}{\mu \theta+\omega(1-\theta)}\mathbb{1}_m\hd{\bm I_{b_c}}
 =\eta_1+\eta_2,$
where
    $\eta_1 = \frac{\left(1-\theta\right)\omega V_{rat}}{\mu \theta+\omega(1-\theta)}\mathbb{1}_m^T\bar{\bm Y}_{22}|\bar{\bm Y}\mathbb{1}_m,
    \eta_2  =\frac{\theta\mathbb{1}_m^T\hd{\bm I^*_c}}{\mu \theta+\omega(1-\theta)}+\frac{\left(1-\theta\right)\mathbb{1}_m^T\bm Y_{12}\bar{\bm Y}_{22}^{-1}\hd{\bm I^*_o}}{\mu \theta+\omega(1-\theta)}.
    $
Since $\theta\in[0,1),$ $\mathbb{1}_m^T\bar{\bm Y}_{22}|\bar{\bm Y}\mathbb{1}_m>0$ and $ \mu>0$, $\eta_1$ is monotonically increasing on $\omega\in\mathcal{W}$. Moreover, since $\mathbb{1}_m^T\bm Y_{12}\bar{\bm Y}_{22}^{-1}\hd{\bm I^*_o}<0$, $\eta_2$ is monotonically increasing on $\omega\in\mathcal{W}$ if $\theta\mathbb{1}_m^T\hd{\bm I^*_c}+\left(1-\theta\right)\mathbb{1}_m^T\bm Y_{12}\bar{\bm Y}_{22}^{-1}\hd{\bm I^*_o}<0$.
Thus, the monotonic increasing  of $\mathbb{1}^T_m\hd{\bm I_c}$ is established.

\bibliographystyle{elsarticle-harv}
\bibliography{hdbai}           

\end{document}